\documentclass[aps,prb,twocolumn]{revtex4}
\def\figwidth{4.2cm}
\def\figwidthw{6cm}
\def\VED{VED\ }

\usepackage{graphics,amsmath}

\begin{document}

\title{On the interface polaron formation in organic field-effect transistors}
\author{G.~De~Filippis$^1$, V.~Cataudella$^1$, S.~Fratini$^{2}$ and S.~Ciuchi$^{3}$}
\affiliation{$^1$CNR-SPIN and Dip. di Scienze Fisiche - Universit\`{a} di Napoli
Federico II - I-80126 Napoli, Italy \\
$^2$Institut N\'eel - CNRS and Universit\'e Joseph Fourier BP 166, F-38042 Grenoble Cedex 9, France\\
$^3$CNR-ISC, CNISM and
Dipartimento di Fisica, Universit\`a dell'Aquila,
via Vetoio, I-67010 Coppito-L'Aquila, Italy}


\begin{abstract}
A model describing the low density carrier state in an
organic single crystal FET with high-$\kappa$ gate dielectrics is studied.
The interplay between  charge carrier coupling 
with inter-molecular vibrations in the bulk of the organic material and 
the long-range interaction induced at the interface with a polar
dielectric is investigated.  This interplay 
is responsible for the stabilization of a  
polaronic state with an internal structure extending on few lattice
sites, at much lower coupling strengths than expected from the polar
interaction alone.
This effect could give rise to polaron self-trapping in high-$\kappa$ organic
FET's without invoking unphysically large values of the carrier interface
interaction.
\end{abstract}

\maketitle
%
%
%


\section{Introduction}

Organic field-effect transistors (OFETs) are the elementary building blocks
of ``plastic electronics''.\cite{GershRMPColl06} In these devices, 
charges are induced in a thin conducting channel at the interface
between an organic semiconductor and a gate
insulator. While commercial applications have
already appeared, 
mostly based on low-cost organic thin-films,
the fundamental physical mechanisms governing the charge dynamics in
these devices are not fully understood. 
As it has become clear in recent years, even when high purity 
crystalline semiconductors are used in order to minimize
the effects of structural and chemical disorder, 
the electronic characteristics of  OFETs are
strongly affected by interactions taking place both 
within the semiconducting material as well as with its close environment.
This can be traced back to the extremely narrow electronic bands 
resulting from the weak van der Waals intermolecular bonding, 
that make organic semiconductors  much more sensitive to
electronic interactions  than  their inorganic counterparts. 
For this reason, polaronic effects have been shown to play an
important role in these devices.  

Several electron-lattice interaction mechanisms  relevant to
organic molecular crystals have been identified and studied in the
past. These include Holstein-like couplings to 
intra-molecular vibrations, leading to a local  reorganization of the
molecular energy levels\cite{HolsteinAnnPhys59I,HolsteinAnnPhys59II,SilbeyJCP80,KenkrePRL89,Hannewald2004} 
as well as
Su-Schrieffer-Heeger (SSH) or Peierls-like 
couplings 
where the transfer integrals between the
molecules are modulated by the inter-molecular
motion.\cite{Friedman,Hannewald2004,SumiSSHJCP79,DukeSchein80, TroisiPRL06}
The latter mechanism couples directly  the electronic motion 
to  the strong lattice fluctuations 
arising from the mechanical softness of these compounds, 
and has been recently 
identified as the dominant limiting factor of  
the charge mobility in pure crystalline
semiconductors. \cite{TroisiPRL06,TroisiAdvMat2007,Zuppiroli2007,FC,Hatch2010}

%
%
In addition to such intrinsic mechanisms, interface-related effects 
have also been demonstrated in 
several systematic studies of organic FETs using 
different
gate dielectrics. 
Sources of charge trapping have been identified, 
either related to the interface
quality\cite{Veres2003,jurchescuPRB09} or to the long-range forces 
arising from polar impurities at the organic/dielectric interface.
\cite{Veres2004,richards08,ZuppiroliPRB10,CramerPRB2010} 
Furthermore, it has been shown 
that the long-range polarization induced 
in high-$\kappa$ gate materials by the charge carriers themselves 
can lead to self-trapping of the carriers.
Even when sufficient care is taken in order to minimize
extrinsic sources of disorder, 
Fr\"ohlich polarons are formed due to the polar long-range interaction of the
carriers in the organic semiconductor  with the phonons at the
semiconductor/dielectric interface. 
The strength of this remote electron-phonon (e-ph) interaction can
be tuned by an appropriate choice of the gate dielectric. The 
``metallic-like'' carrier mobility
characteristic of crystalline organic semiconductors can be converted 
into a much lower,
thermally activated mobility.\cite{HuleaNatMat2006,FratNJP2006,CiukFratPRB2009}

While there is a vast theoretical 
literature dedicated to the different
electron-lattice interaction mechanisms  mentioned
above, 
up to now there have been no attempts to  study the interplay between
bulk and interface e-ph couplings. 
To fill this gap, in this work we analyze a model that includes both  
bulk and interface effects as can be realized in
organic FETs, and  treat them in a common framework.
We focus on the combined effects of the SSH 
coupling of the electronic transfer integrals  to  the
inter-molecular vibrations and of a Fr\"ohlich  long-range e-ph
interaction as induced by the presence of the gate
dielectric in a field-effect geometry.   
Apart from its relevance to organic/dielectric interfaces, from the
fundamental viewpoint 
this model presents an interesting combination of two {\it a priori}
competitive mechanisms.
Our results  show that  
a rather weak SSH coupling strength
as estimated in Refs.[\onlinecite{TroisiPRL06,TroisiAdvMat2007}] 
can have an unexpectedly strong effect in stabilizing 
a  polaronic state when a moderate long-range polar coupling is present. 
Therefore, self-trapped states of small radius can exist 
at common organic/dielectric interfaces,  even in  such cases
where the carrier-interface phonon interaction
alone would not be  sufficient to produce polaronic self-localization 
\cite{ZuppiroliPRB10}. This provides 
a microscopic theoretical basis for
the experimental results of Ref. \onlinecite{HuleaNatMat2006}, where
a finite activation energy indicative of self-trapped states
was observed using gate materials such as   Al$_2$O$_3$ and Ta$_2$O$_5$.

This paper is organized as follows. 
The model under study is introduced in Section \ref{sec:Model}. 
The two methods of solution that will be used 
are described in Section \ref{sec:Methods}.  
In Section \ref{sec:Results} we introduce the main
quantities of interest and present the results of our calculations.  
Section \ref{sec:Conclusions} is devoted to the conclusions.
A detailed derivation of the form of the long-range interaction with the
gate material is
presented in Appendix \ref{sec:Appendix}. 


\section{The model}

We consider the following one dimensional tight-binding model
\label{sec:Model}
\begin{eqnarray}
  H&=& - t\sum_i [1+\alpha (a_i+a^\dagger_i-a_{i+1}-a^\dagger_{i+1})] \; (c^\dagger_{i} c_{i+1} + H.c.) \nonumber\\
  &+&\sum_{i,q} c^\dagger_i c_i (M_q e^{iqR_i}b_q+h.c.) \nonumber\\
  &+&\omega_{SSH}\sum_i a^\dagger_i a_i + \omega_{LR}\sum_q b^\dagger_{q} b_{q}  
  \label{eq:SSH}
\end{eqnarray}
where electrons or holes described by the 
creation and destruction operators $ c^\dagger_{i}, c_{i}$ move on a
lattice labelled by site index $i$. These  interact with molecular displacements  $X_i=a^\dagger_i+a_{i}$ via a SSH 
interaction,
describing 
the transfer integral modulation on the distance between nearest
neighbors with strength $\alpha$, 
as well as with optical modes   
$Y_q=b_{q}+b^\dagger_{-q}$ at the polar interface via a coupling
$M_q$.
To keep the discussion as simple as possible and illustrate the main
consequences of the interplay between the ``off-diagonal'' short range
SSH coupling and  the ``diagonal'' long-range polar coupling, we 
shall restrict the electronic motion to a one-dimensional
chain. Yet, our results can be considered as being qualitatively
representative of organic/dielectric interfaces 
in the more realistic two-dimensional case. 

As it is discussed in Appendix A we consider the one dimensional interface 
model introduced in Refs.[\onlinecite{Stroscio89,Stroscio91}]. 
The e-ph interaction matrix element reads in this case
\begin{eqnarray}
\label{eq:Mq}
M_q=\frac{g}{\sqrt{N}}\sum_i e^{iqR_i} \frac{R^2_0}{R^2_0+R^2_i}
\end{eqnarray}
with $R_0$ a cut-off length of the order of the lattice spacing $a$
and $q$ a strictly one-dimensional vector.
As it is clear from Eq. (\ref{eq:Mq}), the interaction 
between the electron charge and the associated
lattice polarization is long ranged, with a power law decay in real space.
We define two dimensionless coupling constants for the two different 
el-ph couplings: 
\begin{equation}
\lambda_{LR}=\sum_q \frac{|M_q|^2}{2\omega_{LR}t}
\end{equation}
and 
\begin{equation}
\lambda_{SSH}=\frac{\alpha^2 t}{\omega_{SSH}}.
\end{equation}
%
%

We shall study the system in the range of  parameters typical of
single crystal organic semiconductors, taking rubrene as a case study.
We set $\omega_{LR}=0.25t$ as an average value for the interface phonons 
in the case of common polarizable gate materials\cite{HuleaNatMat2006} and 
$\omega_{SSH}=0.05t$ for the relevant intermolecular
vibrations\cite{TroisiAdvMat2007,Hatch2010} ($\hbar=1$ throughout the
paper). The lattice cutoff parameter 
 entering in Eq. (\ref{eq:Mq}) is chosen to be $R_0=0.5a$. 
\cite{FratMorpCiuJPCS2008}  

%
%
%
%


In view of the small values of $\omega_{LR}$ and $\omega_{SSH}$, we will study the system 
by using two approaches: the adiabatic approximation, valid for vanishing $\omega_{LR}$ and $\omega_{SSH}$, 
and a variational exact diagonalization (VED) method able to include non adiabatic contributions.

\section{Methods}
\label{sec:Methods}

\subsection{Adiabatic approximation}
\label{section:ADMethod}

For values of $\omega_{LR}$ and $\omega_{SSH}$ that 
are much smaller than $t$, the lattice
polarization cannot follow the electronic oscillations 
and the wave function of the
system can be factorized\cite{Born} into a product of
normalized variational functions
$\left |\varphi \right\rangle$ and $\left |f \right\rangle$ depending on the electron and phonon coordinates
respectively:
\begin{equation}
\left | \psi \right\rangle =\left | \varphi \right\rangle \left |f \right\rangle
\label{ad1}
\end{equation}
where
\begin{equation}
\left | \varphi \right\rangle =\sum_{R_m}c^{\dagger}_m \left | 0 \right\rangle_{el} \phi(R_m)
\label{ad2}
\end{equation}
This factorization becomes exact when we neglect completely the ionic kinetic energy i.e. $\omega_{SSH},\omega_{LR}\rightarrow 0$ . 
In  Eq.(\ref{ad2}), 
$\left |0 \right\rangle_{el}$ is the electron vacuum state and  $\phi(R_m)$ are
parameters that satisfy the relation:
\begin{equation}
\sum_{m}|\phi(R_m)|^2=1~.
\label{ad3}
\end{equation}

The expectation value of the Hamiltonian (\ref{eq:SSH}) on the state $\left | \varphi \right\rangle $ provides
an effective Hamiltonian for the lattice fields:
\begin{eqnarray}
&&\left \langle \varphi | H | \varphi \right \rangle  =-t\sum_{m} \left [\phi^*(R_m)
\phi(R_m+1)+h.c. \right] \nonumber \\
&& + \sum_{q} \left[
\omega_{LR} b^{\dagger}_{q} b_{q}+ \left( \rho_{q} b_{q}+H.c.\right)
\right]\nonumber \\
&&  +\sum_{m} \left[
\omega_{SSH} a^{\dagger}_{m} a_{m}+ \left( v_{m} a_{m}+H.c.\right)
\right]
\label{ad4}
\end{eqnarray}
with
\begin{equation}
\rho_{q}=M_{q}\sum_{i}e^{i q  R_i}|\phi(R_i)
|^2
\label{ad5}
\end{equation}
and
\begin{equation}
v_m=- \alpha t \left [
  \left ( \phi(R_{m+1})-\phi(R_{m-1}) \right ) \phi^*(R_{m})+h.c. \right]. 
\label{ad6}
\end{equation}

It can be proven that the ground state of this Hamiltonian is a product of two coherent states:
\begin{equation}
\left | f \right \rangle =e^{\sum_{q}\left( \frac{\rho_q}{\omega_{LR}}b_q-h.c.\right)} \left | 0 \right\rangle_{LR}
e^{\sum_{m}\left( \frac{v_m}{\omega_{SSH}}a_m-h.c.\right)}\left | 0 \right\rangle_{SSH},
\label{ad7}
\end{equation}
where $\left | 0 \right\rangle_{LR}$ and  $\left |0 \right\rangle_{SSH}$ are the vacuum states of the two lattice fields. 
We stress that the parameters $v_m$ are real [see Eq.(\ref{ad6})].

At this stage, the phonon 
state $\left | f \right\rangle $ can be used to obtain 
an effective Hamiltonian for the charge carrier:
\begin{eqnarray}
  H_{eff}&=& - t\sum_i [1+2 \alpha (v_i-v_{i+1} )] \; (c^{\dagger}_i  c_{i+1} + c^{\dagger}_{i+1}  c_{i}) \nonumber\\
  &-&\sum_{i} c^{\dagger}_i c_i \sum_{q} (M_q e^{iqR_i} \rho^*_{q}+h.c.)+K 
  \label{elec}
\end{eqnarray}
where 
\begin{equation}
K = \omega_{LR}\sum_{q} |\rho_q|^2 +\omega_{SSH} \sum_{m} v^2_m ~.
\label{ad8}
\end{equation}

The present adiabatic approximation breaks translational
invariance. The problem is reduced to 
studying an electron, confined  by the potential well raised by the LR
interaction. The hopping amplitude from site to site within the well
depends on the specific site pair, via the SSH
coupling.  
The wavefunction 
parameters $\phi(R_m)$ are determined variationally.

\subsection{The \VED diagonalization technique}
\label{section:VEDMethod}

In the systems on which we are focusing our attention the 
phonon frequencies $\omega_{SSH}$ and $\omega_{LR}$ are small compared to the
transfer integrals.\cite{TroisiAdvMat2007}  
The adiabatic approach discussed above should therefore provide a reasonably 
appropriate description of the ground state properties. 
It is known, however, that a proper inclusion of non adiabatic terms
in the  calculation removes the fictitious translational symmetry 
breaking, which is found within the pure adiabatic approximation, 
and can introduce significative changes at any non-zero phonon frequencies. 
Furthermore physical 
quantities related to the quantum nature of the phonons
(number of phonons in the ground state) or the delocalization 
of the electronic wave function (effective mass and ground state
spectral weight) can be accessed only via a non adiabatic approach.
To this aim we implement here a method, combining a variational
approach and an exact diagonalization technique,  
which is able to provide an accurate description 
of the ground state properties 
at all values of the system parameters.
We shall first analyze the two interactions separately,  and then
introduce the method for 
treating the full Hamiltonian Eq. (\ref{eq:SSH}).   

It has been shown\cite{Cat,Cat1,Zhao1,Squeezing} that the ground state
features in the case of ``diagonal'' electron-lattice interactions, 
such as the LR term Eq. (\ref{eq:Mq}),
are very  well reproduced by methods assuming as starting point the approach
introduced by 
Toyozawa.\cite{Toyozawa} 
The idea is to restore a translationally invariant
Bloch state  using the wavefunctions provided by the adiabatic
approximation, i.e.  by taking a superposition of 
localized states centered on different lattice sites. 
The trial wave function that
accounts for the translational symmetry reads:
\begin{equation}
|\psi^{(LR)}_{k} \rangle =\frac{1}{\sqrt{N}}\sum_{n}e^{i k R_n} \left | \psi^{(LR)}_{k}(R_n) \right\rangle
\label{1na}
\end{equation}
where
\begin{equation}
\left | \psi^{(LR)}_{k}(R_n) \right\rangle
=\sum_{l}c^{\dagger}_{l+n}|0\rangle_{el} \;
\phi_{k}(R_l) \left | e^{(LR)}_n \right\rangle
\label{2na}
\end{equation}
and
\begin{equation}
\left | e^{(LR)}_n \right\rangle=
e^{\sum_{q}\left(e^{i q R_n} \frac{\rho_q}{\omega_{LR}}b_q-h.c.\right)} \left | 0 \right\rangle_{LR}.
\label{3na}
\end{equation}
In Eqs.(\ref{1na}-\ref{3na}) the index $n$ denotes the  distance between 
the electron position and the lattice modes. 

The next step is to generalize this wavefunction  
to correctly recover the weak coupling regime where  
the electron is weakly dressed by
lattice oscillations, as predicted by perturbation theory. 
This can be achieved by treating
the parameters entering into the
definition of the function $\rho_q$ as independent variational
parameters,\cite{Zhao1} rather than being constrained by
 the shape of the electron function as in
Eq. (\ref{ad5}). Actually, also in the perturbative regime  the lattice displacement is 
described by a coherent state, but with a different function $\rho_q$ 
including an electron recoil term due to the scattering with phonons.\cite{Cat,Cat1} 
The parameters
 $\phi_k(R_l)$ have to be determined via numerical diagonalization of
 the corresponding Hamiltonian, 
defining the Variational Exact Diagonalization method described
below.  
In this way we obtain a wavefunction that is 
able to capture the weak and strong coupling limits 
and interpolates between them in the intermediate regime.

In practice we use a basis where
each vector is a linear superposition, with appropriate phase factors, of the 
translational copies of a state having a given electron position and
lattice configurations 
(phonons and electron are translated together, with their relative
position being held fixed). 
We choose as phonon states those appearing in Eq. (\ref{3na}) where  
$\rho_q$ is a variational function.
There are then $M$ basis vectors,  
labelled by the $n$ index appearing in Eq. (\ref{3na}), and each of
them includes the variational function $\rho_q$. 
The coefficients $\phi_{k}(R_l)$ in
Eq.(\ref{2na})
are determined 
for each fixed set of the parameters defining the variational function
$\rho_q$ by diagonalizing the Hamiltonian matrix
in the chosen basis.  These do not 
coincide, in general, with the coefficients of the function $\rho_q$ in Eq.(\ref{ad5}).
Summarizing, we have to diagonalize a $MxM$ matrix for each value of the variational function $\rho_q$. 
Finally, 
in the following, we assume that 
the variational parameters $\phi(R_l)$, entering into the definition of the function $\rho_q$ in Eq. 
(\ref{ad5}), 
are different from zero 
up to 3rd nearest neighbors. 
We have checked that this assumption allows to obtain convergency for all the 
values investigated of the model parameters.  

The   procedure described above could 
be straightforwardly generalized to include also the interaction with 
SSH phonons.
Unfortunately, it can be shown that it does not capture the
correct physical features of the SSH Hamiltonian  in the absence of LR
interaction.\cite{Def} 
To treat the SSH term we therefore apply a different
approach based on an exact diagonalization procedure, 
where we do not assume as basis set the coherent states of the adiabatic approach. 
In this approach the phonon variables are efficiently limited to a new set which
is able to describe successfully the ground state properties, as
it will be shown below. 
The real bottleneck comes from the phonon
Hilbert space, that is infinite.
To circumvent this difficulty, we keep 
states where the lattice deformations involve only the sites 
up to 3rd nearest neighbors of the 
lattice site where electron is located, all the others being undeformed
\begin{eqnarray}
\left | ph \right\rangle ({n_j},j=-3,...3) = \prod_{j, i \ne j}^{j=-3,...3} 
\frac {(a_{n+j}^{\dagger})^{n_j} \left | 0 \right\rangle_{n+j}}
{\sqrt{n_j !}} \left| 0\right\rangle_{i}
\;
\label{ph}
\end{eqnarray}
where $n$ indicates the electron position. We apply 
a truncation procedure considering only the states with at most $L$ quanta, 
i.e. $\sum_j n_j \le L$, where 
$L$ is fixed by requiring convergency of the ground state energy. 
These states provide a correct physical description of the ground state 
properties, since the lattice deformations are effectively 
localized around the site where the electron is located. 
Furthermore, in analogy with
the LR coupling, we use
the translational symmetry associated to
periodic boundary conditions to reduce the size of the sparse
Hamiltonian matrix, requiring that the states have a definite momentum.

When both LR and SSH couplings are considered, we adopt a 
\VED diagonalization 
technique on a truncated phonon basis that combines the two subspaces 
introduced separately for the 
$\alpha=0$ and $g=0$ cases. This basis is given by the product of the
vectors describing the two different couplings  
separately, Eqs. (\ref{3na}) and (\ref{ph}). The ground state is found by using 
the modified Lanczos algorithm.\cite{Dagotto}
We emphasize that this 
approach retains the full translational invariance requiring that
each basis vector has a well defined momentum. 

\section{Results}
\label{sec:Results}

%

\begin{figure}[!h]
\centering
\resizebox{\figwidth}{!}{\includegraphics{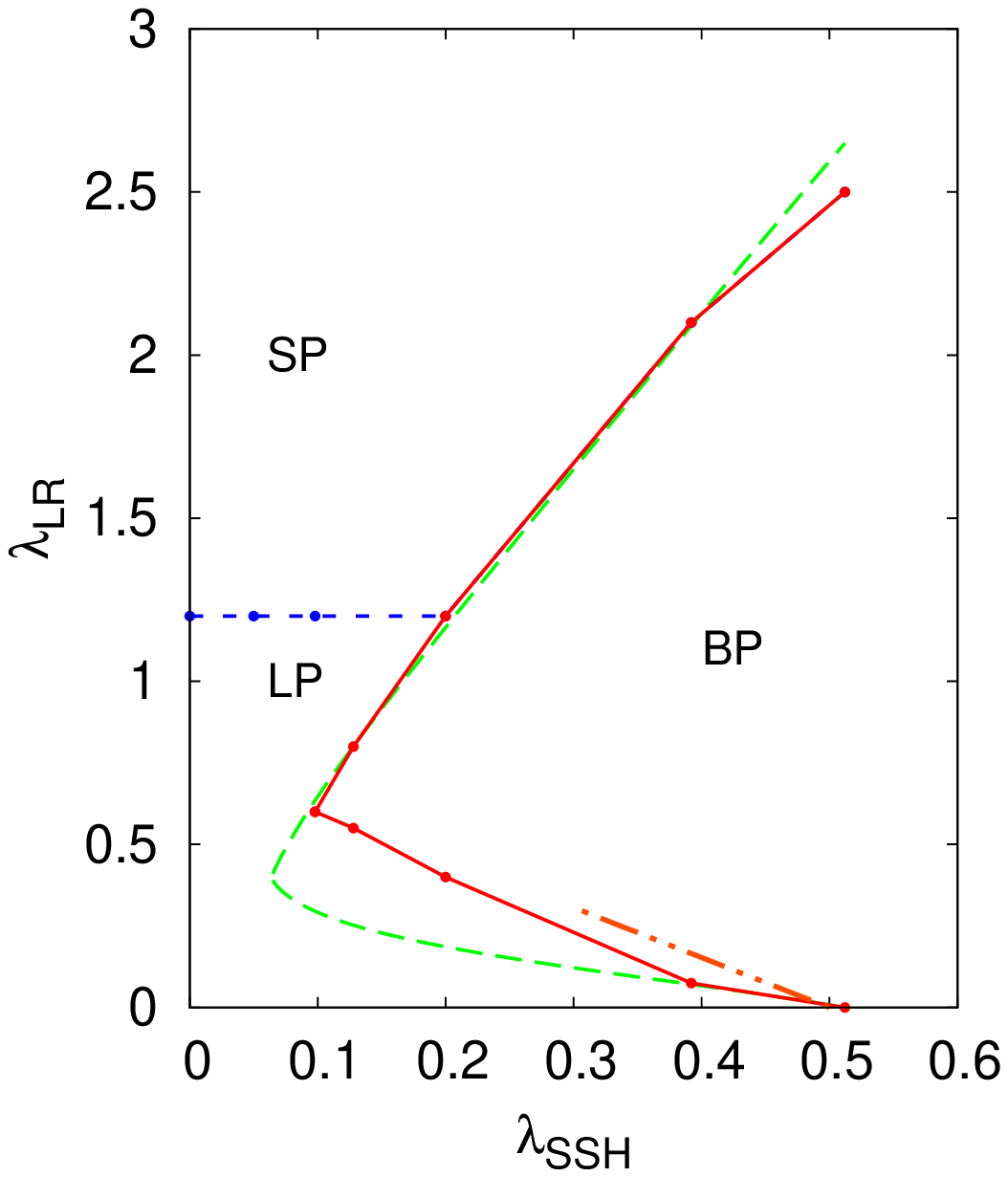}}
\resizebox{\figwidth}{!}{\includegraphics{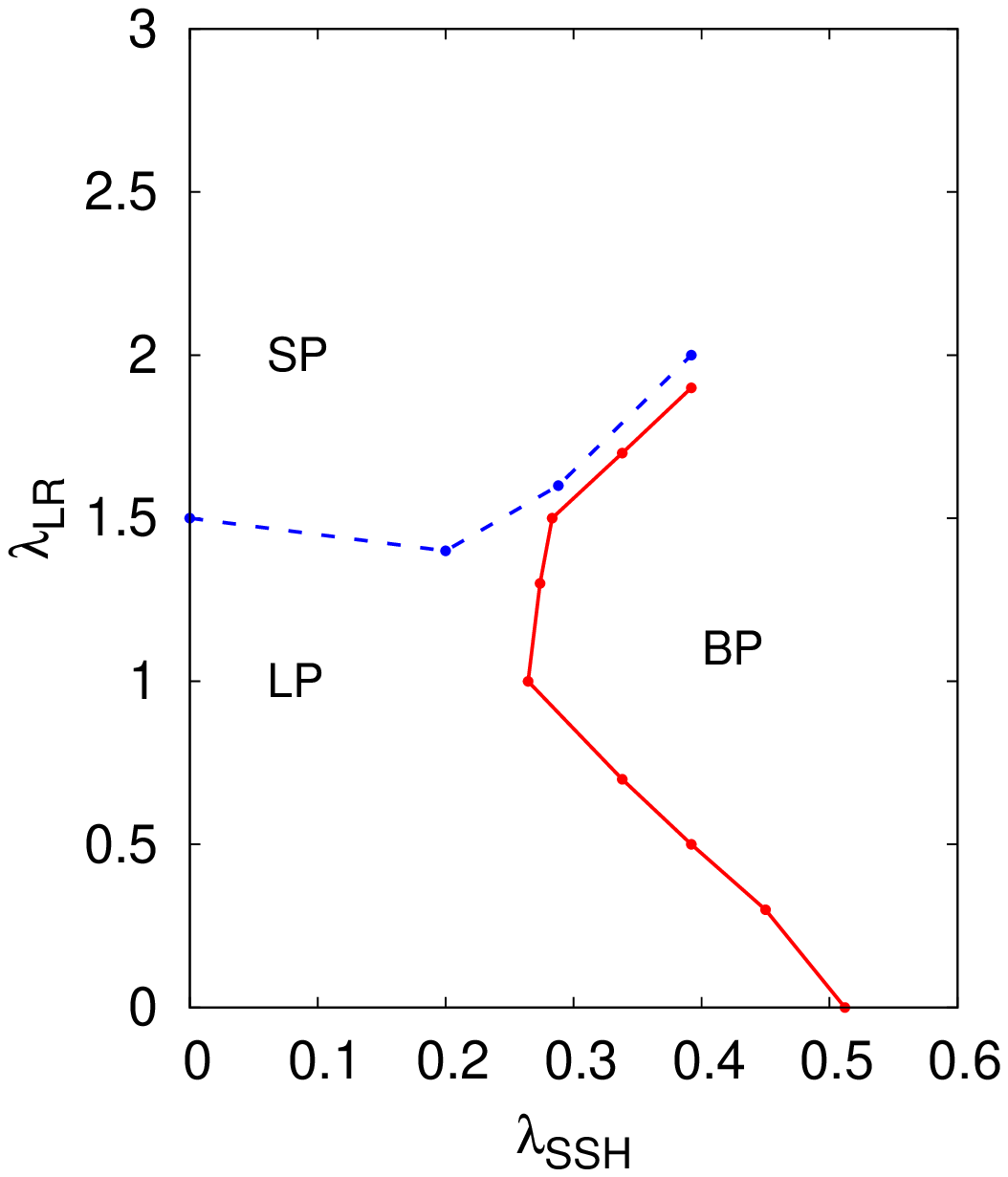}}
\caption{Phase diagram a) in the adiabatic approximation and b) in the
  \VED approximation.
The dashed (blue) line marks 
the crossover from large (LP) to small (SP) site-centered polaron 
probed by the bimodality of the LR-PDF. 
The solid (red) line indicates the crossover to the bond polaron (BP) state
probed by the bimodality 
of the SSH-PDF while the dashed (green) line is the transition to the
bond polaron state  probed by the electronic wavefunction (only in
panel a).  
In panel a) the dashed double dotted red line is the
analytic estimate of Eq. (\ref{eq:reentr}).
}
\label{fig:phasediag}
\end{figure}
Due to their different microscopic nature, the 
two interaction terms in Eq. (1) induce different types
of polaron states: the LR coupling to the electron density favors 
localized states centered on the lattice sites (site polaron), 
while the SSH coupling to the electronic transfer integrals
tends to localize the polarons on
inter-molecular bonds, i.e. with the electronic charge 
centered between two neighboring molecular sites (bond polaron).
This competition results in the phase diagrams reported in  
Fig. \ref{fig:phasediag}, which constitute the central results of this
paper. The most notable feature is the existence of a
reentrant phase where a bond polaron state is stabilized by the
interplay between the two microscopic interaction mechanisms. 
This result is robust
 against the presence of phonon quantum fluctuations, as is apparent from the
 comparison of the phase diagram obtained from the adiabatic
 approximation,  valid in the limit of vanishing
 frequencies $\omega_{SSH},\omega_{LR}\rightarrow 0$
 [Fig.\ref{fig:phasediag} (a)], with the 
 results at finite phonon frequencies [Fig.\ref{fig:phasediag} (b)]. 


\subsection{Adiabatic approximation}
\label{section:res_ad}

In the adiabatic approximation the ground state electronic wave function is
always localized. 
However, for small values of $\lambda_{SSH}$, a crossover  
from a large to a small site-centered polaron occurs upon 
increasing the LR interaction strength $\lambda_{LR}$ 
[blue dashed line in Fig. \ref{fig:phasediag} (a)]:
the wavefunction extension is progressively reduced, 
being always centered on a given molecular site.
The locus of the crossover can be determined by analyzing the qualitative 
changes induced on the phonon wavefunction, as described at the end of
this Section.
%
%
The large to small polaron crossover
is only weakly affected by the presence of the SSH 
interaction.

\begin{figure}[!h]
\centering
 \resizebox{\figwidthw}{!}{\includegraphics{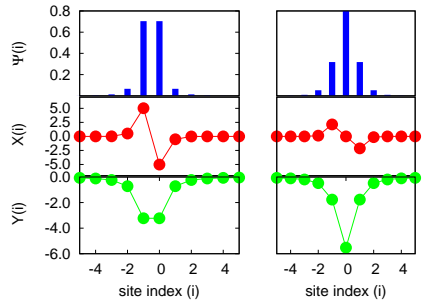}}\\
\caption{Bond polaron properties for $\lambda_{SSH}=0.4$
$\lambda_{LR}=1.0$ (left panels) and $\lambda_{LR}=1.5$ (right panels) 
in the adiabatic approximation. 
Upper panel: the adiabatic electronic probability.  
Central panel: SSH average distortions of the organic molecular lattice. 
Lower panel: LR average distortions at the organic/dielectric
interface.}
\label{fig:adiabatic-wf}
\end{figure}

As $\lambda_{SSH}$ increases, 
the symmetry of the electronic wave function
changes continuously from site-centered to bond-centered.
The breaking of translational invariance allows us to analyze this
transition by direct inspection of the electronic wavefunction, 
which is shown in Fig. \ref{fig:adiabatic-wf} 
together with its associated distortions in the bond-centered (left panel)
and site-centered polaron (right panel) states.
The transition between the two phases is 
indicated by the green dashed line in Fig.\ref{fig:phasediag}a.
%
%
%

Contrary to the large/small polaron crossover, that is rather
insensitive to the value of $\lambda_{SSH}$, the site to bond-polaron
transition is   significantly modified by the presence of the
 LR interaction, which
strongly stabilizes the bond polaron state.  The boundary of the 
bond-centered polaron phase shows a markedly reentrant behavior, 
reaching a minimum $\lambda_{SSH}=0.072$ at $\lambda_{LR}=0.4$. 
\cite{noteCapone}

\bigskip

It is instructive to understand the 
stabilization of the bond polaron phase induced by the 
long-range interface interactions from 
the following energy balance argument. In the absence of intermolecular
coupling ($\lambda_{SSH}=0$) the energy of the large polaron can
be written in general as
\begin{equation}
E_{LP}=-2t-\Delta E_{LP}
\end{equation}
where the second term can be expanded 
for small $\lambda_{LR}$
as $\Delta E_{LP}=t(a_1\lambda_{LR}+a_2\lambda^2_{LR}+...)$. 
On the other hand we see from Fig. \ref{fig:adiabatic-wf}
that in the bond polaron phase the
adiabatic electronic wave function is mainly localized on two
neighboring sites, with exponentially vanishing tails on the adjacent
molecules.  
Neglecting such tails provides the following estimate  for
the bond polaron energy:
\begin{equation} 
E_{BP}=-t(1+2\lambda_{SSH})-t(1+\eta)\lambda_{LR}
\end{equation}
where the two contributions 
come respectively from SSH and LR interactions.
The parameter $\eta=\sum_q |M_q|^2 e^{iaq}/\sum_q |M_q|^2$ ($a$
the lattice spacing, $0<\eta<1$) 
takes into account the spatial variation of the
long-range potential-well between two neighboring sites. Comparing the
two energies we obtain  a transition  when
\begin{equation}
\label{eq:reentr}
\lambda_{SSH}=1/2+[a_1-(1+\eta)/2]\lambda_{LR},
\end{equation}
and a reentrant behavior is found when $a_1<(1+\eta)/2$. Observing that
$\eta$ is largest for slowly varying potentials, we recognize
that the reentrance is strongly favored by 
the long range nature of the polar dielectric interaction. 
The reason is that such long-ranged
polar interaction, for which the preferred
polaronic state would be site-centered, is 
not extremely sensitive to the local details of the wavefunction, so that
it can still provide a sizable energy stabilization 
in the case of a dimer structure. 
It should be stressed that the
arguments presented here are only qualitative. For a quantitative
analysis of the large polaron to bond polaron 
phase boundary, the extended nature of the bond polaron wavefunction
must be taken into account, which further enhances the reentrant
behavior as compared to Eq. (\ref{eq:reentr}) (see
Fig.\ref{fig:phasediag}). 

\bigskip

Phonon related characteristics also change across the bond polaron
transition, following the changes of the electronic wavefunction. 
As  can be seen from Fig. \ref{fig:adiabatic-wf}, 
the average SSH distortions within a bond polaron  are
opposite, since this configuration minimizes the SSH 
interaction energy. 
Within the entire 
bond polaron region of the phase diagram,  
even at the relatively large value of the LR coupling $\lambda_{LR}=1$
illustrated in Fig. \ref{fig:adiabatic-wf}, the LR distortions
follow the charge distribution determined by minimizing the SSH
interaction energy.  

The phonon probability distribution functions (PDF) provide an alternative
tool to characterize the ground state properties,\cite{MaxCiukPiovraEPL98} 
with the advantage of 
being directly generalizable to the translationally
invariant ground states examined in the next section. 
The SSH-PDF, which represents the  distribution of SSH displacements at
site $i+j$ given an 
electron on site $i$, is defined as
\begin{equation}
P^{(j)}_{SSH}=\left< |X_{i+j}>c^\dagger_i |0> <0| c_i <X_{i+j}| \right>
\label{eq:PDF}
\end{equation}
where $|X_{i+j}>$ is the coordinate eigenstate corresponding to the
displacement of SSH phonons on site 
$i+j$. The LR-PDF is defined by substituting $X$ with $Y$ in Eq. (\ref{eq:PDF}).
It is worth noting that the 
PDF is closely related to the distributions of interatomic distances that can be
directly measured by neutron spectroscopy such as
pulsed neutron diffraction.\cite{EgamiPRB97}

\begin{figure}[!h]
\centering
  \resizebox{\figwidth}{!}{\includegraphics{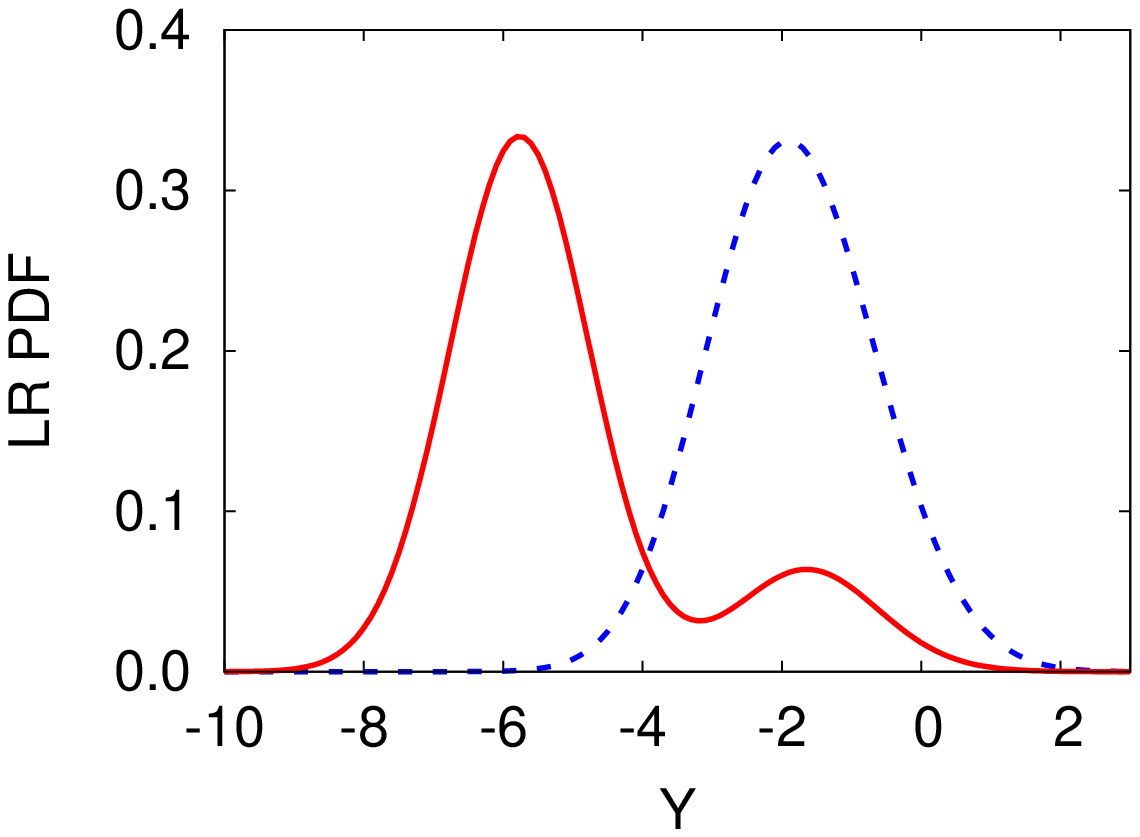}}
  \resizebox{\figwidth}{!}{\includegraphics{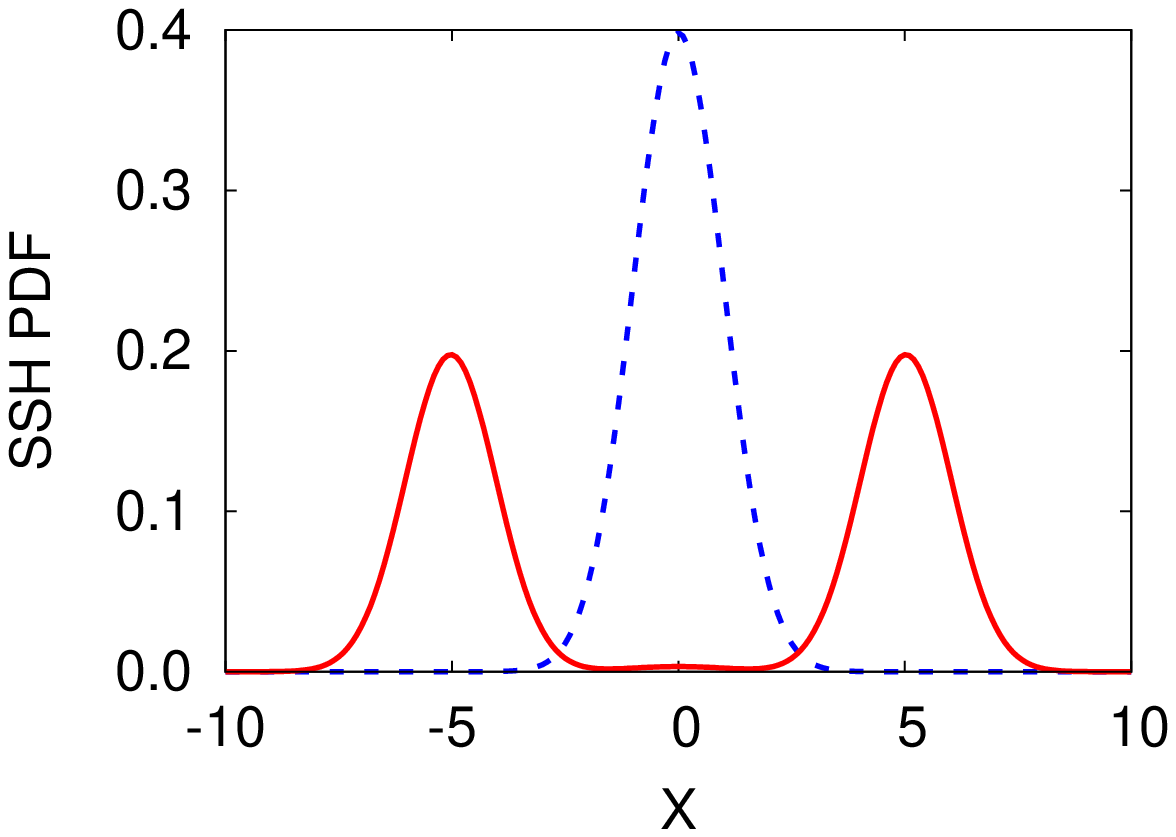}}\\
\caption{(left) Local probability distribution function
  $P^{(0)}_{LR}(Y)$ of the LR phonons
 (see text) at 
$\lambda_{SSH}=0$,
$\lambda_{LR}=0.7$ (blue dotted line) and $\lambda_{LR}=1.5$ (red solid line);
(right) Same for the SSH phonons: $P^{(0)}_{SSH}(X)$ at 
$\lambda_{SSH}=0.4$,
$\lambda_{LR}=0$ (blue dotted line) and $\lambda_{LR}=1$ (red solid line)
}
\label{fig:adiabatic-ppdf}
\end{figure}

The local ($j=0$) part of these functions is plotted
in Fig. \ref{fig:adiabatic-ppdf} for different polaronic ground states. 
In the small 
$\lambda_{SSH}$
region of the phase diagram [see
Fig. \ref{fig:phasediag} a)], the large-to-small polaron
crossover  is highlighted by the bimodality in the LR-PDF [panel
(a) of Fig. \ref{fig:adiabatic-ppdf}]. On the other hand, when the 
bond polaron state is {\it well} formed, the SSH-PDF shows two symmetric peaks
[panel (b)], whose distance increases with 
$\lambda_{SSH}$.
This provides an alternative method to locate the bond polaron region,
which is shown in Fig. \ref{fig:phasediag} (a) as a red continuous
line. The resulting phase diagram shows that a broad
crossover region emerges (between the green and red lines in
Fig. \ref{fig:phasediag} where the electronic wavefunction has already
undergone the bond transition but a real bimodality of the SSH-PDF is
still not observed, with the SSH-PDF showing large non-Gaussian fluctuations.
Let us note that in the bond polaron region of the
phase diagram, the analysis of the LR-PDF does not show any 
 evidence of a large to small bond polaron
crossover, at least for  values of
$\lambda_{SSH} \le 0.5$. 

Summarizing,  the adiabatic phase diagram shows three distinct regimes: 
small site-centered polaron, large site-centered polaron and 
bond polaron.  
The bond polaron regime, with the electron wavefunction mostly
localized on a molecular dimer, is stabilized by the presence of even 
a moderate long range coupling. 

\subsection{Variational Exact Diagonalization method}

Fig. \ref{fig:phasediag} (b) illustrates the phase diagram
obtained 
using the \VED scheme described in section \ref{section:VEDMethod}. 
Translational symmetry breaking is prevented 
by the quantum  fluctuations of both SSH and LR phonons.
Accordingly, all the lines shown in Fig. \ref{fig:phasediag} (b)
correspond to {\it crossover} lines separating translationally
invariant states with 
different characteristics, as obtained by  analyzing the 
 phonon PDFs. \cite{MaxCiukPiovraEPL98}


\begin{figure}[!h]
\centering
  \resizebox{\figwidth}{!}{\includegraphics{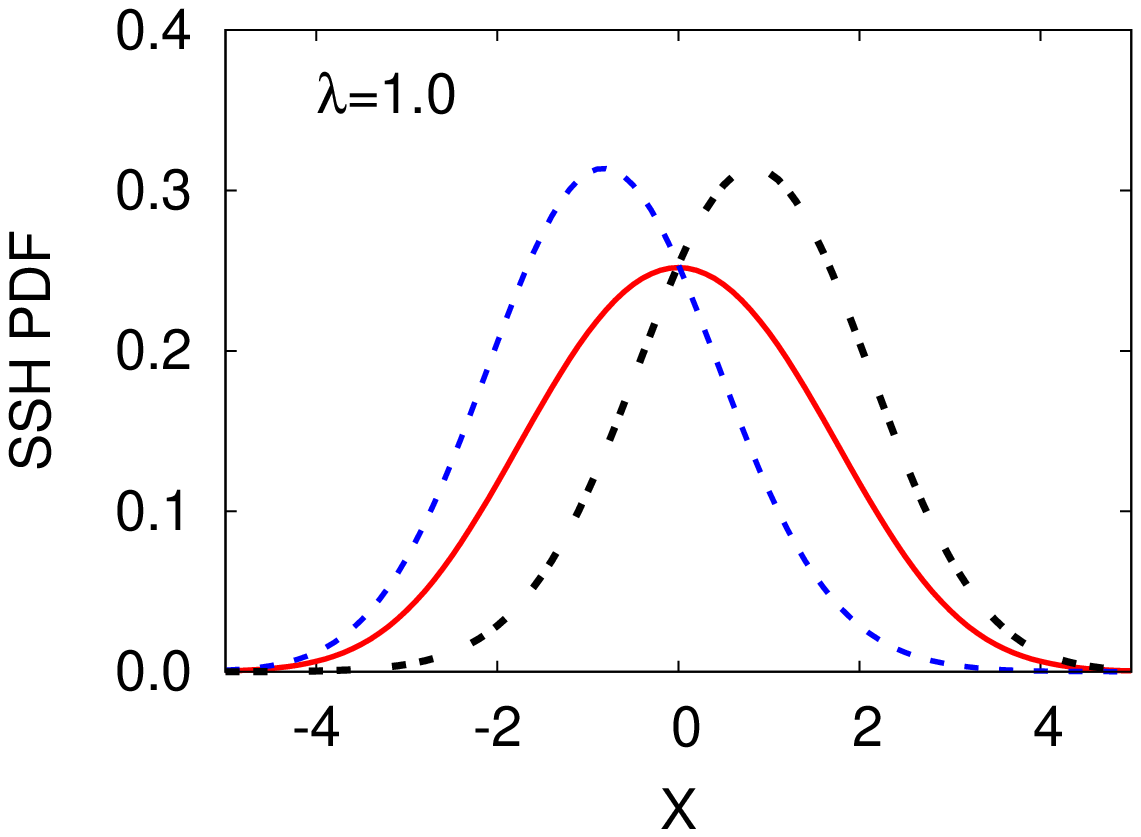}}
  \resizebox{\figwidth}{!}{\includegraphics{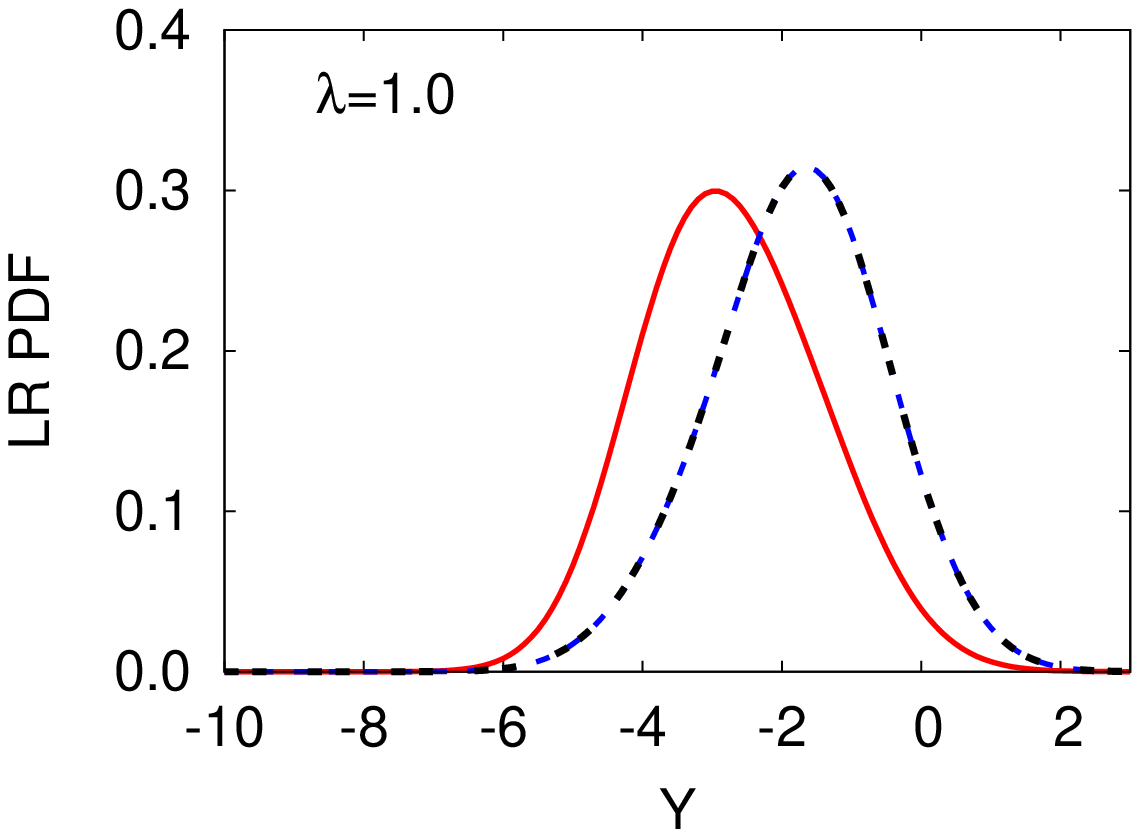}}\\
  \resizebox{\figwidth}{!}{\includegraphics{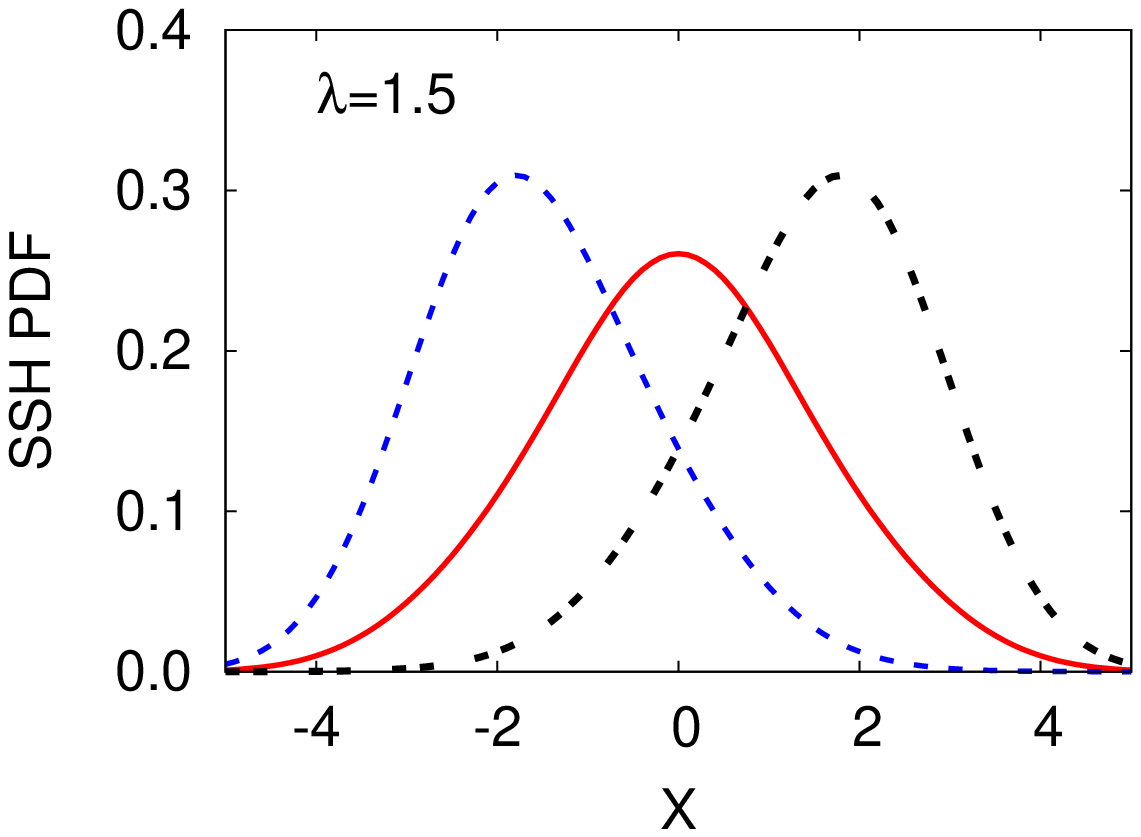}}
  \resizebox{\figwidth}{!}{\includegraphics{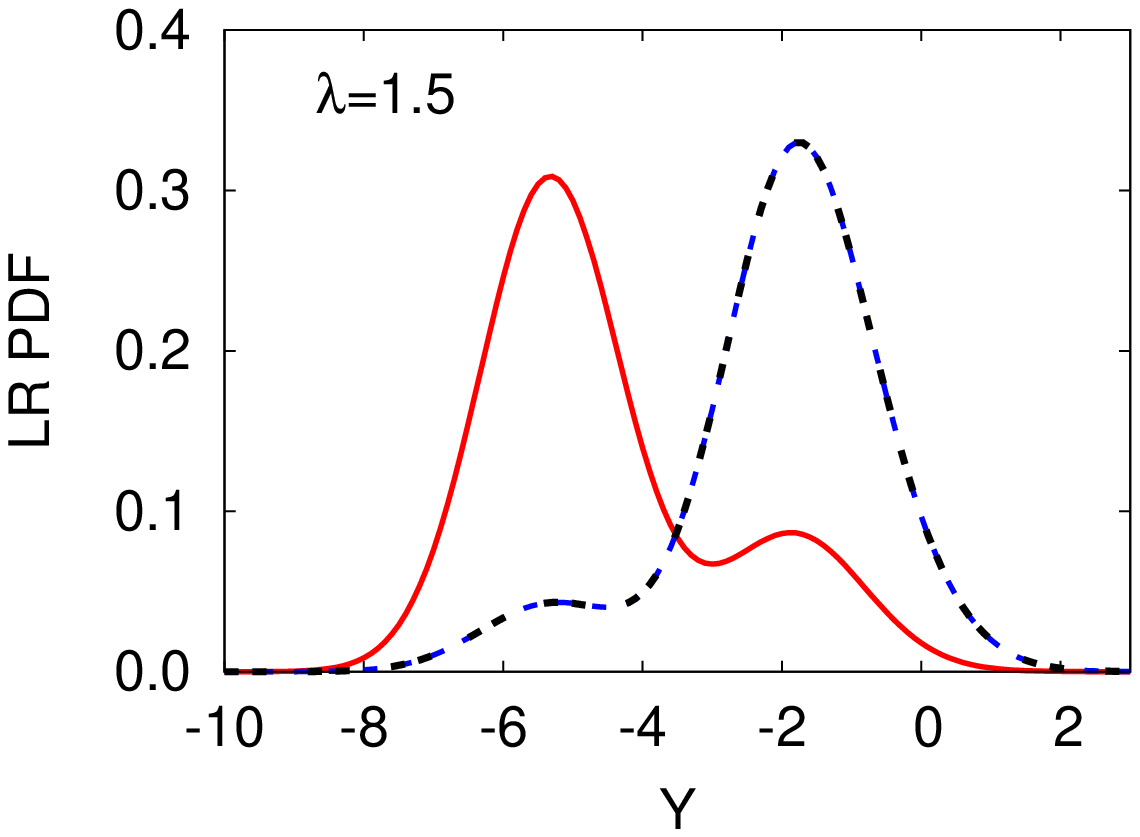}}\\
  \resizebox{\figwidth}{!}{\includegraphics{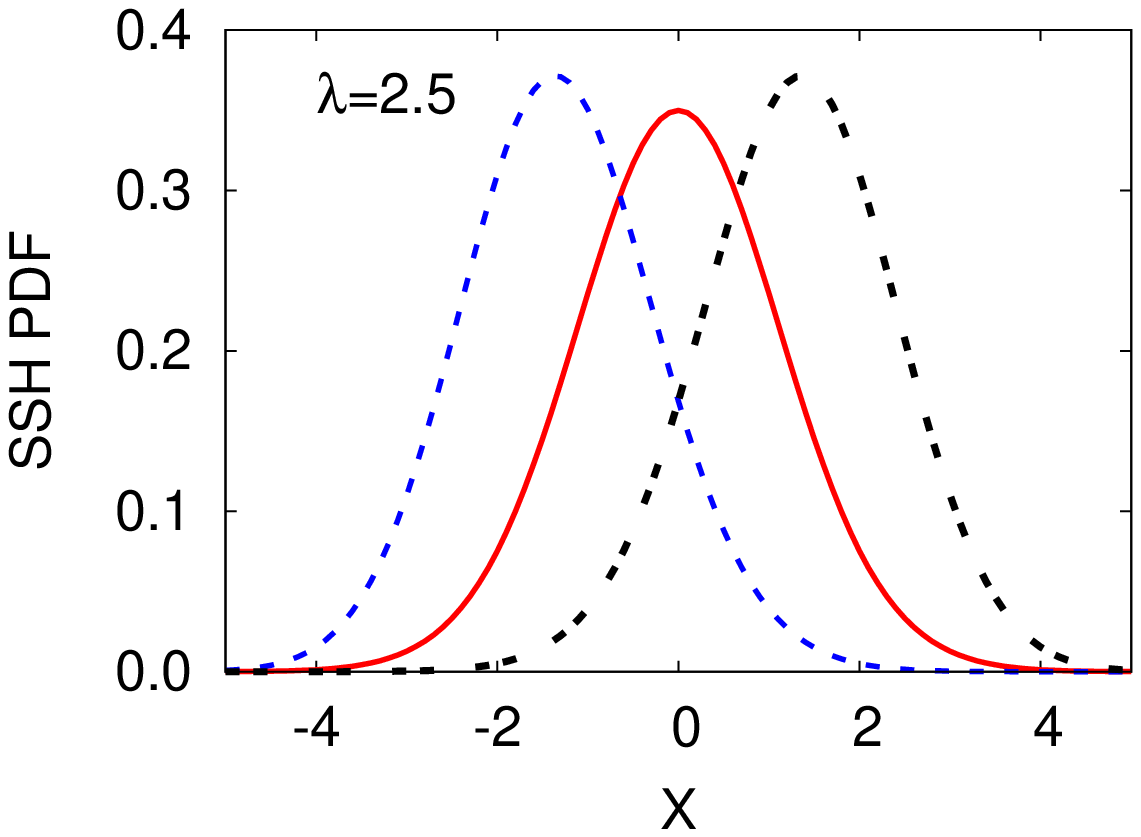}}
  \resizebox{\figwidth}{!}{\includegraphics{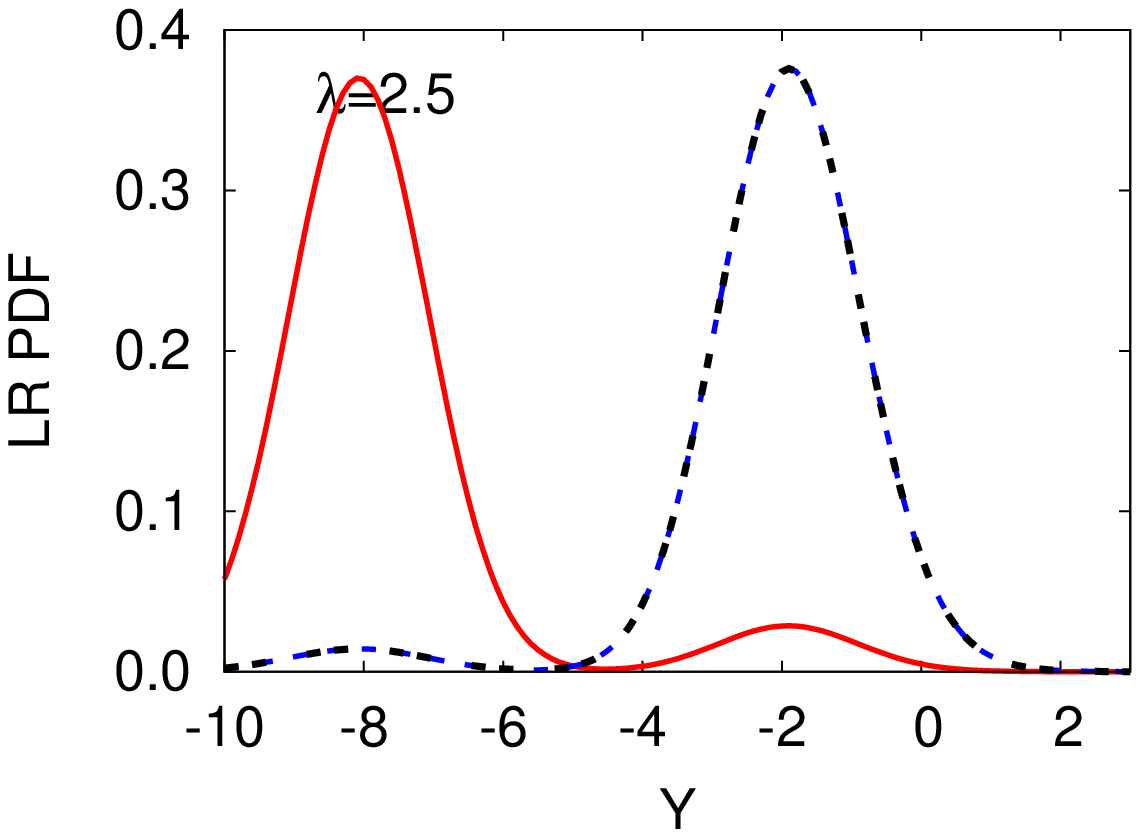}}
\caption{SSH (left) and LR (right): phonon PDF at 
$\lambda_{SSH}=0.2$: upper panel: $\lambda_{LR}=1.0$, central panel: $\lambda_{LR}=1.5$, 
lower panel: $\lambda_{LR}=2.5$. Solid (red) line represents $j=0$ (see text); dashed blue (double dashed black) 
line represents $j=1$ ($j=-1$).}
\label{fig:pdfnormal}
\end{figure}
In Fig. \ref{fig:pdfnormal} 
we plot the SSH-PDF and LR-PDF for $j=-1,0,1$, in the small
$\lambda_{SSH}$ region (we use $\lambda_{SSH}=0.2$)
of the phase diagram, across
the large-to-small polaron  crossover  [see Fig. \ref{fig:phasediag} (b)]. 
This crossover is evinced by the
bimodality in the LR-PDF, occurring 
 at $\lambda_{LR} \simeq 1.4$. 
Both the on-site ($j=0$) 
and the inter-site  LR-PDF ($j=-1,1$, which are
equal by symmetry), become bimodal at the crossover.  
In this region, the SSH phonons present  a nearly Gaussian
distribution with opposite distortions around the  
polaron localization site. It is interesting to note that the average value
of the SSH displacement has a non-monotonic dependence on the LR 
interaction and it is largest for intermediate values of
$\lambda_{LR}$, signaling the proximity to the 
bond polaron regime.

\begin{figure}[!h]
\centering
  \resizebox{\figwidth}{!}{\includegraphics{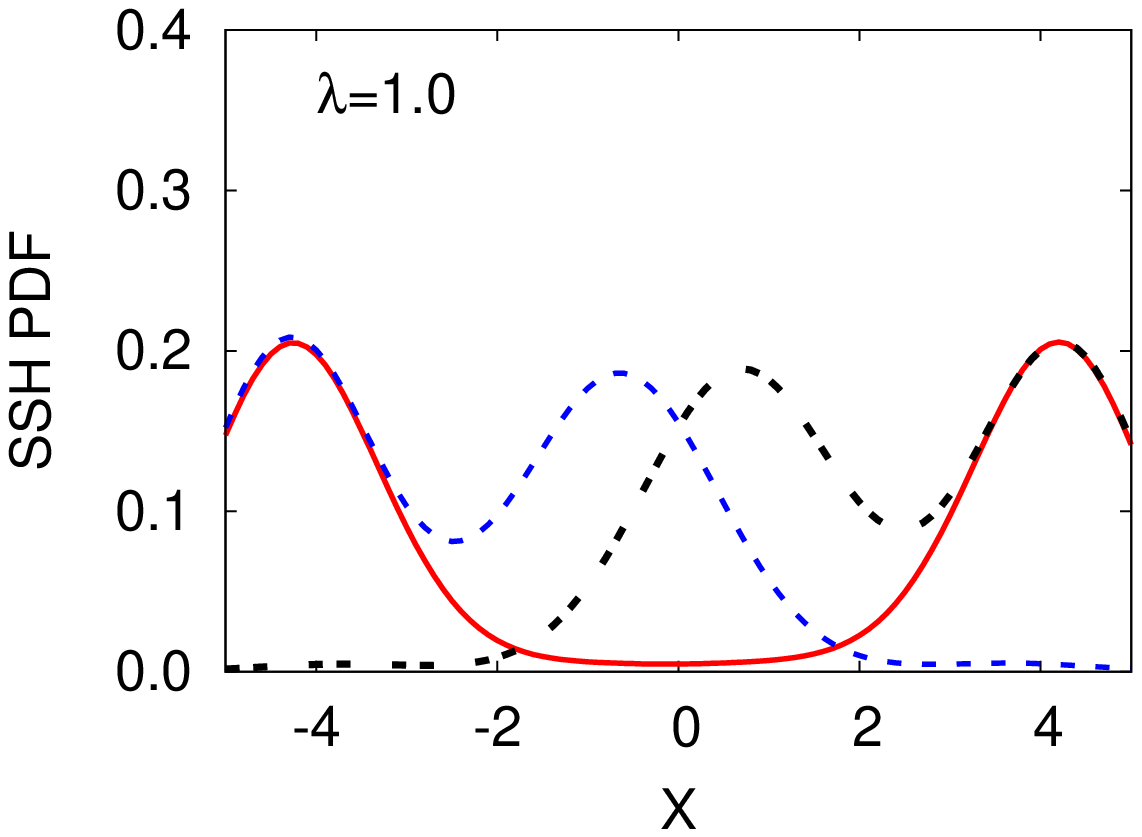}}
  \resizebox{\figwidth}{!}{\includegraphics{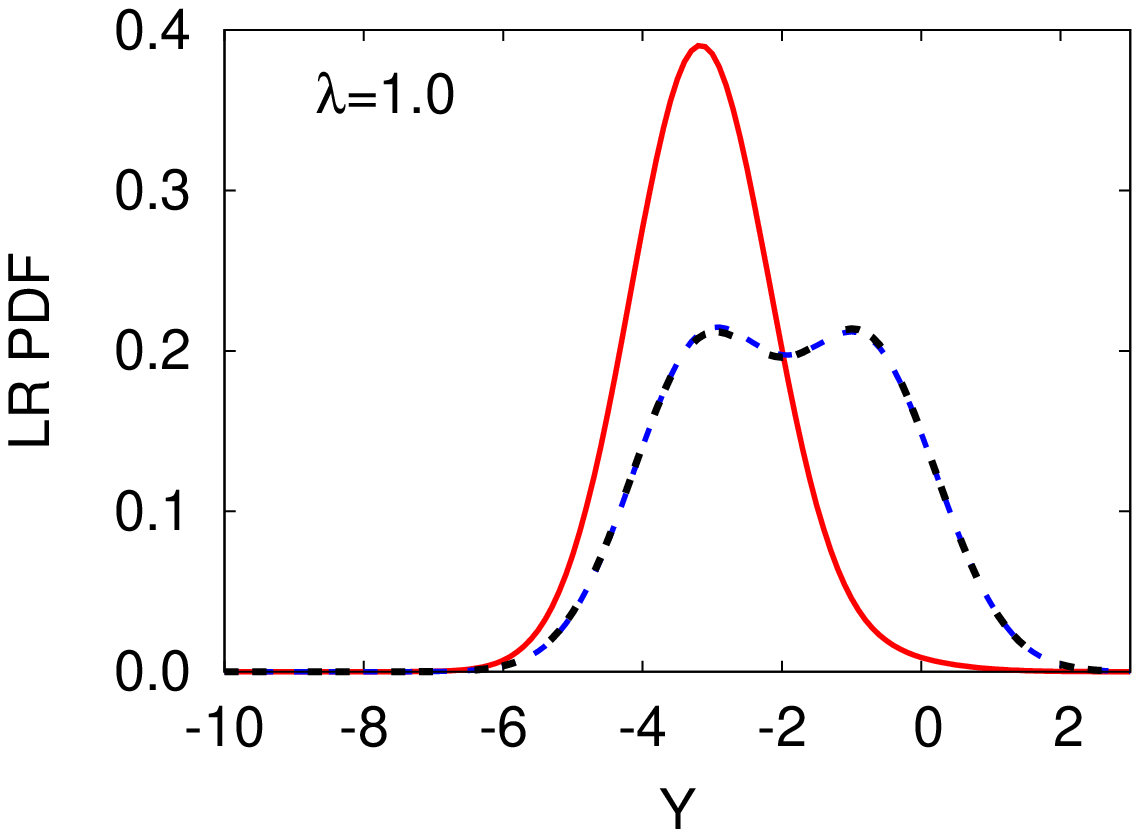}}\\
\caption{ 
SSH (left) and LR (right) phonon PDF at 
$\lambda_{SSH}=0.4$ in the bond polaron regime. 
The solid (red) line represents the on-site contribution, $j=0$ (see
text); dashed blue (double dashed black)
line represents the inter-site part $j=1$ ($j=-1$).}
\label{fig:pdfintermediate}
\end{figure}
In the bond polaron phase both the
LR and the SSH-PDF have a bimodal character, as shown in
Fig. \ref{fig:pdfintermediate}. 
The SSH-PDF displays bimodality in both the on-site ($P^{(0)}_{SSH}$) and
nearest neighbor ($P^{(\pm 1)}_{SSH}$) contributions. 
The site where the polaron is located belongs to either a left or 
right-centered dimer, a situation which is consistent  
with the translational invariance of the ground state. 
The LR-PDF instead is bimodal only
on the nearest neighbor  
($P^{(\pm 1)}_{LR}$), with two peaks corresponding to
distorted and non-distorted sites respectively. The presence  
of a distortion with the same amplitude on nearest neighboring sites is
another signal of the tendency to form a dimer. This tendency is also 
displayed by the LR
distortions    which,
in this region of the phase space,  follow the charge
distribution dictated by the SSH interaction. 

\begin{figure}[!h]
\centering
\resizebox{\figwidthw}{!}{\includegraphics{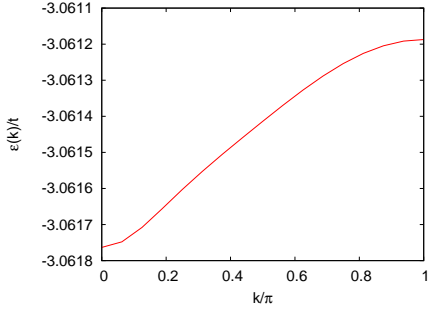}}
\caption{
Polaron band at $\lambda_{LR}=1.0$ and $\lambda_{SSH}=0.288$ (reentrant bond polaron phase), 
in units of $t$.}
\label{fig:band}
\end{figure}

Let us note that 
the reentrant 
bond polaron phase at $\lambda_{LR}\ne 0$ has a
 different symmetry than  that at $\lambda_{LR}=0$.
Indeed, according to Ref.[\onlinecite{ZhaoJChemPhys08}], 
the transition to the bond polaron phase induces the ground state
wavenumber to change continuously from $k=0$ to $k=\pm \pi/2$ in a
narrow region around 
$\lambda_{SSH}=0.5$.
As soon as $\lambda_{LR}$ is different from zero, we do not find any
evidence of a $k\ne 0$ ground state for $\lambda_{SSH}<0.5$. 
To illustrate this point, 
the electron dispersion characteristic of this phase, where 
the minimum is located at $k=0$, is plotted in Fig. \ref{fig:band}. 
We conclude that in the reentrant bond polaron phase,
the LR coupling stabilizes a Bloch wavefunction having $k=0$.    

\begin{figure}[!h]
\centering
  \resizebox{\figwidthw}{!}{\includegraphics{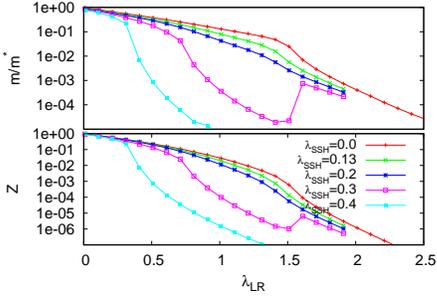}}
\caption{
Inverse effective mass renormalization (upper panel, 
$m$ is the  band mass) 
and the quasiparticle spectral weight (lower panel)
as a function of $\lambda_{LR}$ for different values of $\lambda_{SSH}$.}
\label{fig:massvsLambda}
\end{figure}
In Fig. \ref{fig:massvsLambda} we plot the inverse 
effective mass and the quasiparticle weight, 
defined as $Z=\left|\left \langle GS \right| c^\dagger_{k=0} |0 \right \rangle |^2$,
as a function  of $\lambda_{LR}$ for different values 
of the SSH coupling $\lambda_{SSH}$. 
At small vaues of $\lambda_{SSH}$ the inverse effective mass
first  decreases smoothly with $\lambda_{LR}$ and then 
drops at the large-to-small polaron
crossover around $\lambda_{LR} \simeq 1.4-1.5$. The quasiparticle spectral weight shows a similar
dependence on $\lambda_{LR}$, although with a steeper reduction 
which is characteristic of non-local e-ph interactions.\cite{alex,wellein} 
In contrast with the rather smooth $\lambda_{LR}$ dependence,
a much stronger dependence on $\lambda_{SSH}$ can be
seen in Fig. \ref{fig:massvsLambda}, signaling the crossover into the
reentrant bond polaron regime at $\lambda_{SSH}<0.5$.
We note that the quasiparticle spectral weight
and inverse effective mass have in this case 
the same behavior with $\lambda_{LR}$ as it occurs for local e-ph
interactions, \cite{wellein} i.e. they tend to be mutually proportional.

\begin{figure}[!h]
\centering
  \resizebox{\figwidthw}{!}{\includegraphics{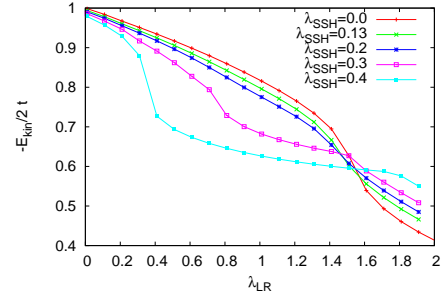}}
\caption{Electron kinetic energy, in units of the bare kinetic energy 
($\lambda_{SSH}=\lambda_{LR}=0$), 
as a function of $\lambda_{LR}$ for different values of
$\lambda_{SSH}$.}
\label{fig:EkinvsLambda}
\end{figure}

Complementary insight on the electron dynamics in the different phases
can be obtained from the analysis of  the total kinetic energy,
$E_{kin}=-t\sum_i \left < (c^{\dagger}_i  c_{i+1} +
  c^{\dagger}_{i+1}  c_{i}) \right >$,
shown in Fig. \ref{fig:EkinvsLambda}. Unlike the quasiparticle spectral weight
$Z$ and effective mass $m^*$,   this quantity gets contributions
from both  the coherent long distance motion of the polaron, and
from the internal motion of the electron within the lattice potential-well.
The large-to-small polaron crossover at small $\lambda_{SSH}$ causes 
a sharp drop of $E_{kin}$ around $\lambda_{LR}\simeq 1.5$. 
For 
$\lambda_{SSH}>0.25$
a region emerges where the kinetic energy stays
relatively constant over a broad range of $\lambda_{LR}$. 
This plateau is another clear signature of the  
bond polaron phase. Here the effective mass and  the
quasiparticle spectral weight are very small, indicative of
polaron self-trapping,  yet a large contribution to the
kinetic energy arises from the internal dimer structure of the  
polaron, with the electron moving freely along the intermolecular
localization bond. 

\begin{figure}[!h]
\centering
  \resizebox{\figwidthw}{!}{\includegraphics{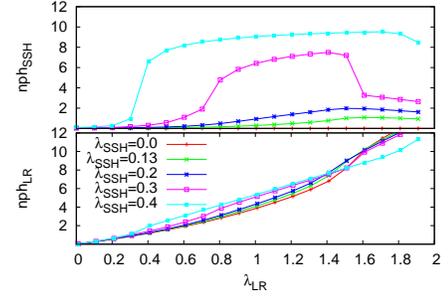}}
\caption{Average number of SSH (upper panel) and LR phonons (lower
  panel) in the polaronic ground state
as function of $\lambda_{LR}$ for 
several values of $\lambda_{SSH}$.}
\label{fig:nphvsLambda}
\end{figure}
The large number of SSH phonons participating in the polaronic state 
in the reentrant phase, shown in Fig. 
\ref{fig:nphvsLambda}, indicates that a strong coupling with
intermolecular modes in the organic crystal can be effectively 
triggered even  at  relatively weak values of $\lambda_{SSH}$,
by suitably turning on the LR interaction.  
At the same time, the number of LR phonons stays relatively small, 
exhibiting only a  weak dependence on $\lambda_{SSH}$ (see lower panel of
Fig. \ref{fig:nphvsLambda}).


\section{Conclusions}
\label{sec:Conclusions}
In this paper, we have investigated a microscopic 
model that describes the interplay between the inter-molecular 
electron-phonon coupling
and the additional interaction arising
 in crystalline organic semiconductors due to the long range
polarization at the interface with a gate dielectric, two ingredients that
are commonly found in organic field effect transistors. 
The ground state properties obtained by two complementary approaches
--- the adiabatic approximation and a more refined variational
method that correctly preserves translational invariance in the
presence of phonon quantum fluctuations --- 
demonstrate the  synergistic cooperation
between bulk and interfacial interactions.   
In particular, we have found that a large-mass polaronic state with a dimer
structure  and  strong inter-molecular
distance fluctuations is stabilized  at
relatively low values of both the long-range polar coupling
$\lambda_{LR}$ and the intermolecular coupling $\lambda_{SSH}$.
Such {\it bond polaron} reentrant phase arises
in a region of the phase diagram where the two interactions taken separately
would not give rise to strong polaronic effects.  
The long range nature of the polar interaction is
crucial for the establishment of this cooperative interplay. 
The latter occurs because the long range interface polarization is
able to provide a sizable energy gain
without much perturbing the local dimer structure
dictated by the SSH intermolecular coupling, 
so that the system can take advantage of
both microscopic interaction mechanisms simultaneously.


From the device perspective, our findings provide a microscopic basis
to explain the existence of small-radius self-trapped states  without invoking
unrealistic values of the polar coupling.
Indeed, the electrical characteristics of high-$\kappa$
organic FETs have been analyzed in Ref.[\onlinecite{HuleaNatMat2006}] 
in terms of the polar interface interaction alone, and the observed
thermally activated mobilities are not entirely 
consistent with the estimated values of $\lambda_{LR}$, 
as has been pointed out in Ref.[\onlinecite{ZuppiroliPRB10}].
The present results show that the proper inclusion of a bulk
electron-lattice interaction mechanism such as the intermolecular SSH
coupling extends the region of small-radius polaronic self-trapping
down to much lower values of $\lambda_{LR}$.    
Let us stress that an analogous interplay as the one studied here can be
expected when combining the effects of a bulk inter-molecular
electron-lattice coupling and the interaction with localized
polarizable impurities such as those considered in Refs.
[\onlinecite{richards08},\onlinecite{ZuppiroliPRB10}].
%

 
Comparing the  inter-molecular coupling strength estimated for Rubrene,
 $\lambda_{SSH}\simeq 0.17$, 
\cite{TroisiAdvMat2007} with the phase diagrams of
Fig. \ref{fig:phasediag}, 
we conclude that interfaces involving this organic material and the 
aforementioned gate materials lie in a region 
of the phase diagram where the intermolecular/gate
synergistic effects are expected to be of relevance. More importantly, 
our results could explain why
at present there are  only few organic semiconductors (among which,
Rubrene) where metallic-like mobilities have been measured in a FET
configuration. Other materials with slightly larger values of the
inter-molecular coupling strength,  $\lambda_{SSH}$, would
rapidly enter the bond-polaron phase, causing the carriers to undergo a  
self-trapping transition as soon as a moderate
interaction with the dielectric environment is present.


Although the model investigated in this paper includes some drastic
approximations, i.e. the model is one-dimensional and the real lattice
structure of the organic FET is not taken into account, we believe
that the main conclusions provided here will not
be qualitatively modified by the inclusion of more realistic lattice
structures and interactions.
The detailed study of the
transport characteristics in the bond polaron phase identified here
remains as an open question for future work.

\appendix
\section{Long-range interaction at the interface.}
\label{sec:Appendix}

In this work we studied the interplay between SSH and Fr\"ohlich
e-ph interactions in one-dimension.
While the definition of the SSH coupling in one-dimension is
unambiguous,\cite{TroisiAdvMat2007} the polar e-ph interaction term depends
on the geometry of the interface. We now examine two physically different cases.

The case that is studied in the paper is a truly one-dimensional
interface, derived for a wire of square cross-section and permittivity $\epsilon_1$
embedded in a three-dimensional polar dielectric with permittivity $\epsilon_2$.\cite{Stroscio89,Stroscio91} 
We specialize the results of Ref.[\onlinecite{Stroscio91}] by allowing the motion of the charges only
on the symmetry axis of the interface $z$ with $x=0,y=0$,
at a distance $R_0$ from the
polar dielectric with permittivity $\epsilon_2$. The medium with permittivity $\epsilon_1$
is taken to be the vacuum. In this geometry the interaction of electrons
with the interface phonons reads
\begin{equation}
 \label{Stroscio}
  M_{q}= M_0 e^{-qR_0}
\end{equation}
with $R_0=(|x|+|y|)/\sqrt{2}$
or, in real space,
\begin{equation}
\label{eq:1d}
  f(R)\propto \frac{R_0}{R^2+R_0^2}
\end{equation}

Notice that here the phononic modes are purely one dimensional
modes, i.e. they  are propagating waves only
along the wire (direction $z$).


It is interesting to compare the one dimensional interaction discussed above with a two dimensional model.
The model that we consider is the generalization of the case studied
previously by Mori and Ando,\cite{MoriAndoPRB89} that describes the
coupling between electrons and two-dimensional interface phonons. In
this case the interaction matrix element in momentum space reads
$M_q=M_0 e^{-qz}/\sqrt{q}$, with  the distance
$z$ between the electrons  and the interface acting as a short-distance cutoff.
Performing the Fourier transform we find the Hypergeometric function
\begin{equation}
  f(R)=\frac{M_0 \pi^{3/2}}{z^{3/2}} {}_2F_1(3/4,5/4,1,-R^2/z^2)
\end{equation}
This scales as $1/R^{3/2}$
for large $R\gg z$, and tends to a
constant $f(0)=M_0 \sqrt{\pi}/z^{3/2}$ for $R\ll z$. This function is well
approximated by the power-law form
\begin{equation}
  \label{eq:2d}
  f(R)=\frac{ M_0^\prime}{(R^2+R_0^2)^{3/4}}
\end{equation}
where $R_0=(4\sqrt{\pi}c)^{2/3} z\simeq 0.66 z$ and $M_0^\prime=cM_0$.
It can be easily checked by retransforming back to k-space that this
power-law form
gives rise to the correct $q\to 0$ limit and lies within $10\%$ of the
original one for all $q\lesssim 1/R_0$.

Then, in both one and two-dimensional
cases, Eqs. (\ref{eq:1d},\ref{eq:2d}) give rise to  long range interactions,
describing qualitatively similar physical situations.
We have chosen to study explicitly the first model in the body of the
paper, as in that case the  properties of the  phononic degrees of
freedom  can be directly accessed
by the solution of a genuinely one dimensional model.

\section*{References}

\begin{thebibliography}{10}
\bibitem{noteCapone}
We note that the site-polaron to bond-polaron transition
sharpens upon reducing the LR coupling strength, until 
 it approaches $\lambda_{SSH}=0.5$ at $\lambda_{LR}=0$.
This value marks the onset of the "pathological"
localization found in Ref.[\onlinecite{CaponePRB97}] 
in the static phonon limit, where 
the SSH coupling leads to a two-site polaron dimer 
that is perfectly isolated from the rest of the system, due to 
the vanishing of the corresponding transfer integrals. 
The model remains however well-defined even for larger
values $\lambda_{SSH}$, \cite{ZhaoJChemPhys08}
whose analysis 
is  beyond the scope of the present paper.
\end{thebibliography}


\end{document}